# Characterization of researchers in condensed matter physics using simple quantity based on $h$-index

M. A. Pustovoit

*Petersburg Nuclear Physics Institute, Gatchina Leningrad district 188300 Russia*

Analysis of citation records of 52 active and productive condensed matter physicists shows that the ratio of $h$-index to the mean age of $h$ most highly cited publications is a reliable quantity that allows meaningful comparison of scientists of different age.

## Introduction

The need in quantitative estimation of individual researcher's scientific level has greatly increased in last years. This led to numerous attempts to obtain simple and informative indices of scientist's output based on citations of his or her publications. Probably the most interesting and promising example is the recently proposed Hirsch index ($h$-index) [1]. It can be obtained simply by sorting the publications of a researcher (or any relevant group of researchers) in descending order by the number of citations $N_c$ and finding the maximum index $h$ in the list for which $N_c \neq h$. Popular databases like Thomson ISI or Scopus show $h$-index in their citation analysis. Besides of its simplicity, $h$-index is informative being considered together with total number of citations and with scientific age[1]. Using a fairly large sample of prominent physicists, Hirsch made interesting observations (e.g., mean h-index about 40 for Nobel prize winners compared with 45 as a lower threshold for membership in US National Academy of Sciences) and provocative suggestions (e.g., taking a physicist as outstanding when he/she has $h$-index 40 after 20 years of career).

It was pointed out later that $h$-index has imperfections and limitations [2-5]. It is not surprising for a single-number measure, so it would be useful to find other simple quantities that can be combined with $h$-index to obtain a more satisfactory picture of individual scientific activity.

In our study we analyze the large sample of researchers in a rather narrow field (52 condensed matter physicists) in order to obtain a fair statistics for several important quantities. We found that the ratio of $h$-index to the mean age of publications that contribute to $h$-index (that is, $h$ most highly cited publications) is a good quantity for comparison of scientists of different age. We also consider the problem of self-citations and show that the level of self-citations in our sample is low.

## Details of the sample

Selection of condensed matter physicists was made by the following criteria: i) the unique name that gives a single result in ISI author search; ii) the considerable productivity, i.e., ten or larger publications in Physical Review B, Physical Review Letters or arxiv.org/cond-mat as a result of author search at the website; iii) the ongoing scientific activity, that is, at least one publication in the current year (2006) in the sources mentioned above. In addition, care was taken to keep the balance between experimentalists and theorists and between ages. The special attention was given to exclude coauthorship between selected scientists. The final sample included 52 researchers. The following quantities were measured for every of them using Thomson ISI Web of Knowledge:

1) total number $N$ of publications authored (ISI does not account for preprints, conference thesis and book chapters);

2) Total number $C$ of publications that cite the person's publications. This number is definitely less than the total number of citations $N_c$. However, the former can be obtained much easier than the latter;
3) The Hirsch index $h$;
4) The mean age $T_h$ of $h$ most cited papers. The age of a particular publication is the difference between 2006 and the year of this publication.
5) The number $N_s$ of self-citing papers. This quantity is easily obtained by sorting the list of citing papers by author in descending order.
6) The scientific age $T$ of a person defined as the difference between 2006 and the year of the first publication in the database. Since ISI contains records from year 1955, we assume that this quantity does not differ from the true scientific age.

**Index $p$ and its properties**

We construct a new index as a ratio of the Hirsch index and the mean age of papers that contribute to it:

$$p = \frac{h}{T_h} \tag{1}$$

In order to examine its properties, let us consider the "linear citation model" used by Hirsch [1]. In that model the scientist writes $n$ papers every year, and each paper receives $c$ citations every year beginning with the year next to its publication. Noting as $y$ the year of publication of most recent papers those contribute to the $h$-index, we get after $T$ years:

$$ny = h,$$
$$(T - y)c = h. \tag{2}$$

One can easily derive from here both $h$ and $y$. Since

$$T_h \approx T - \frac{y}{2}, \tag{3}$$

we obtain:

$$p = \frac{2nc}{2n + c}. \tag{4}$$

The extreme cases are:

$$p \simeq 2n, \quad c \gg n,$$
$$p \simeq c, \quad c \ll n. \tag{5}$$

In words, the index $p$ for low citation rate is limited by this quantity, and for high one by productivity. This means that its value is unlikely to be very high, and one can expect that for particular field of science the distribution of this index is rather narrow.

The index $p$ might resemble the supplemental index $m$ proposed by Hirsch [1], $m = h/T$. Indeed, for the "linear citation model" we have:

$$m = \frac{nc}{n+c}, \qquad (6)$$

which is very similar to Eq.(4). However, i) $p$ can be obtained from real database records much easier than $m$, and ii) it favors high citation rate:

$$\frac{p}{m} = 1 + \frac{c}{2n+c} \simeq \begin{cases} 2, & c \gg n, \\ 1, & c \ll n. \end{cases} \qquad (7)$$

**Results and discussion**

The quantities of special attention in this paper are the Hirsch index and the mean age of publications that form it. The distributions of $h$ and $T_h$ are shown in Figure 1.

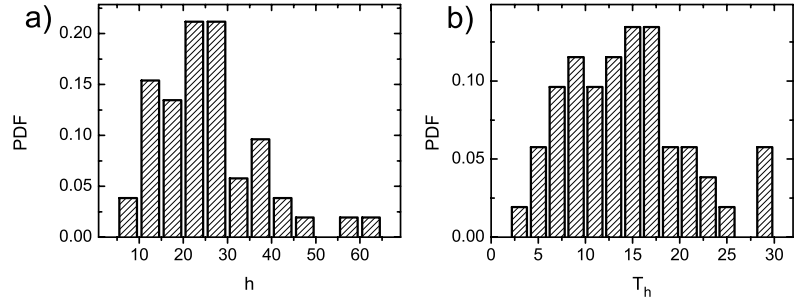

Figure 1. Distribution of our sample of condensed matter scientists: (a) by their Hirsch index, $h$, and (b) by the mean age of $h$ most cited publications, $T_h$. The median values are 24 for $h$ and 14 years for $T_h$.

We see that the distribution of $h$ has a single sharp peak near $h = 25$ followed by a rather long tail. Probably this value can be considered as typical for condensed matter physics, and the tail is formed, in particular, by such persons as J.G.Bednorz ($h = 40$) and J.D.Jorgensen ($h = 63$). The distribution of publication age $T_h$ is single-peaked around 15 years; such behavior probably is a result of our selection procedure.

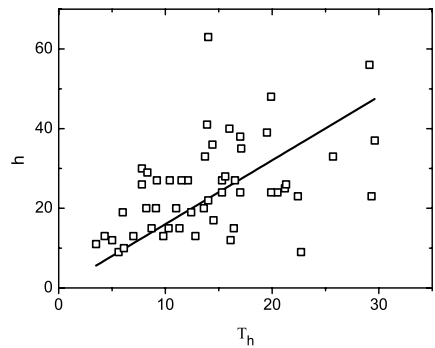

Figure 2. Scatter plot of $h$ versus $T_h$. Line is a one-parameter fit of the data. The correlation coefficient is $r = 0.46$.

Figure 2 demonstrates significant linear correlation between $h$ and $T_h$ (correlation coefficient is $r = 0.46$), and the usage of the new index $p$ as their ratio is well justified here.

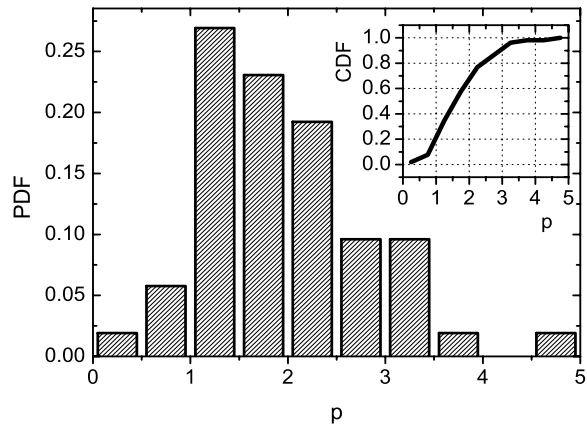

Figure 3. Probability density of the index $p$. The median value is $p = 1.78$. Inset: cumulative distribution function of $p$; only 10% of individuals have $p > 3$.

Figure 3 shows the distribution of $p$. It is again single-peaked around $p = 2$. Hence one can see that a successful condensed matter physicist with $h = 20$ wrote his/her highly cited papers ten years ago on average. The inset of Figure 3 shows the cumulative distribution function (the probability that a scientist's $p$ is less than a certain value). We see that about 90% of considered persons have $p < 3$. Hence we can suggest that an individual with this index higher than 3 is either a respected scientist with large $h$ formed by his/her famous papers written some time ago, or a young active one whose Hirsch index is due to considerable amount of highly cited recent papers. The examination of our sample supports this suggestion, and the numbers of young and well-known scientists appear to be equal. We stress here that "young" should be understood here as "scientifically young", with most of highly cited papers written recently. Such property is one of the main advantages of the index $p$.

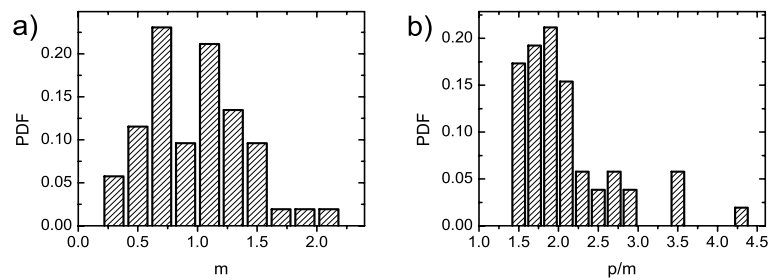

Figure 4. a) Probability density of the index $m$; b) the same for the ratio $p/m$.

For comparison, the distribution of the index $m$ is shown in Figure 4a. It has two peaks, for unclear reasons. It is just interesting to note that most of scientists in our sample have $m$ less than 2, the value suggested by Hirsch [1] for "outstanding" physicist.

The correlation coefficient between $p$ and $m$ in our sample has a high value 0.83. Thus we can safely use the ratio $p/m$ as a reliable index. Its distribution is shown in Figure 4b. The most notable result is the heavy tail with $p/m > 2$ (the maximum value in Eq.(7)), which clearly demonstrates not only the fail of "linear citation model" but the superiority of $p$ over $m$ in

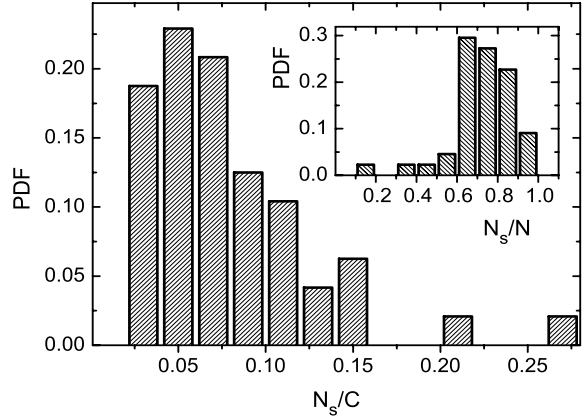

Figure 5. Distributions of fractions of self-citing papers $N_s/C$ (main graph) and $N_s/N$ (inset).

Finally, we estimate the role of self-citations in our sample. Figure 5 displays the distribution of fractions of self-citing papers $N_s/C$ and $N_s/N$. We see that the first one is low (on average, only 8% of citing papers), and the second one is high (which is natural for a scientist involved in a particular problem for some time).

**Summary and conclusions**

We obtain citation data for a representative sample of 52 scientists working in the rather narrow field, the physics of condensed matter. We calculate the mean age $T_h$ of highly cited papers that contribute to individual's Hirsch index $h$. This quantity can be fairly easy obtained from database search results. Its rather strong correlation with $h$-index allows to construct the new index $p = h/T_h$. In our sample this index has a distribution with a single peak around $p = 2$, that is, an average successful condensed matter physicist with Hirsch index $h$ wrote his/her highly cited papers $h/2$ years ago on average. The small fraction (about 10%) of the individuals who have $p > 3$ includes respected as well as promising scientists. The index $p$ is correlated to the quantity $m$, the rate of increase of $h$-index, but is more easily obtainable and favors scientists who wrote their highly cited papers recently.

**Acknowledgements**

The author is grateful for support and hospitality to Laboratory of Physical and Structural Biology, National Institute of Child Health and Human Development (USA), where most of the data has been collected. The work was also partially supported by RFBR grant 05-02-17626 and Russian State programs "Macrophysics" and "Strongly correlated electrons in semiconductors, metals, superconductors and magnetic materials".

**References**
1. J. E. HIRSCH, An index to quantify an individual's scientific output, *Proceedings of the National Academy of Sciences of the United States of America*, 102 (2005) 16569–6572 [http://arxiv.org/abs/physics/0508025].
2. C.D. KELLY, M.D. JENNIONS, The h-index and career assessment by numbers, *Trends in Ecology and Evolution,* 21 (2006) 167-170.
3. M. SCHREIBER, Self-citation corrections for the Hirsch index, *Europhysics Letters*, 78 (2007) 30002 [http://arxiv.org/abs/physics/0701231].
4. B. JIN, L. LIANG, R. ROUSSEAU, L. EGGHE, The *R*- and *AR*-indices: Complementing the *h*-index, *Chinese Science Bulletin,* 52 (6) (2007) 855-863.
5. J. E. HIRSCH, Does the h-index have predictive power?, http://arxiv.org/abs/0708.0646.